\documentclass[12pt]{iopart}

\usepackage{iopams}
\usepackage{graphicx}

\begin{document}

\title[An improved two-way continuous-variable quantum key distribution]{An improved two-way continuous-variable quantum key distribution protocol with added noise in homodyne detection}

\author{Maozhu Sun, Xiang Peng and Hong Guo}

\address{The State Key Laboratory of Advanced Optical Communication Systems and Networks, School of Electronics Engineering and Computer Science, Peking University, Beijing 100871, China}
\ead{\mailto{xiangpeng@pku.edu.cn}, \mailto{hongguo@pku.edu.cn}}

\begin{abstract}
We propose an improved two-way continuous-variable quantum key distribution (CV QKD) protocol by adding proper random noise on the receiver's homodyne detection, the security of which is analysed against general collective attacks. The simulation result under the collective entangling cloner attack indicates that despite the correlation between two-way channels decreases the secret key rate relative to the uncorrelated channels slightly, the performance of the two-way protocol is still far beyond that of the one-way protocols. Importantly, the added noise in detection is beneficial for the secret key rate and the tolerable excess noise of this two-way protocol. With the reasonable reconciliation efficiency of $90\%$, the two-way CV QKD with added noise allows the distribution of secret keys over 60 km fibre distance.

\end{abstract}

\submitto{\jpb}
\maketitle

\section{Introduction}
Quantum key distribution can enable two authentic parties, the sender (Alice) and the receiver (Bob), to obtain unconditional secret keys without restricting the power of the eavesdropper (Eve) \cite{Scarani,BB84}. On the premise of unconditional security, the higher key rate and the longer distance are constantly pursued \cite{postselectionSilberhorn,improvingdistanceALeverrier}. To enhance the tolerable excess noise of the continuous-variable quantum key distribution (CV QKD) \cite{squeezedstateprotocolMHillery,2Grosshans02}, the two-way CV QKD protocols are proposed \cite{6Pirandola,8Pirandola}, where Bob initially sends a mode to Alice, and Alice encodes her information by applying a random displacement operator to the received mode and then sends it back to Bob. Bob detects both his original mode and received mode  to decode Alice's modulations. Although the two-way CV QKD protocols can remarkably enhance the tolerable excess noise \cite{6Pirandola,8Pirandola}, it needs to implement the tomography of the quantum channels to analyze the security under general collective attack \cite{6Pirandola}, which is complicated in practice. Therefore, we proposed a feasible modified two-way protocol by replacing the displacement operation of the original two-way protocol with a passive operation on Alice's side \cite{ourpaper}. However, the source noise \cite{sourcenoiseRFilip,sourcenoiseRFilip2010,sourcenoiseaddednoiseenhancekeyWeedbrook,sourcenoiseYongShen,sourcenoiseYShen} and both detection efficiency and detection noise \cite{detectnoiseJLodewyck,KRB1JLodewyck} on Bob's side are not considered in the modified protocol.

It has been proved that adding a proper noise on Bob's detection side in one-way CV QKD can enhance the tolerable excess noise and the secret key rate in reverse reconciliation \cite{sourcenoiseaddednoiseenhancekeyWeedbrook,8Cerfaddednoiseenhancekey,Weedbrookaddednoiseenhancekey,Pirandolaaddednoiseenhancekey,Renneraddednoiseenhancekey,Renesaddednoiseenhancekey}. This idea has been applied to the original two-way protocol in \cite{MWangandWPan}, while the scheme did not consider the correlation between the two channels. The correlated noise affects the secret key rate \cite{correlatednoiseZengGuihua2012,correlatednoiseZengGuihua2011}. In this paper, we apply the idea of adding noise to our modified two-way protocol $\textrm{Hom}_\textrm{M}^2$ \cite{ourpaper} to enhance the tolerable excess noise and the secret key rate. Considering the correlation between the channels, the security of the two-way CV QKD with added noise against entangling cloner collective attacks \cite{Grosshans03reventanglecloner,CWeedbrookentanglecloner} is analysed and numerically simulated.

\section{The two-way CV QKD with added noise in homodyne detection}

\begin{figure}
\centering
\includegraphics[width=10.5cm]{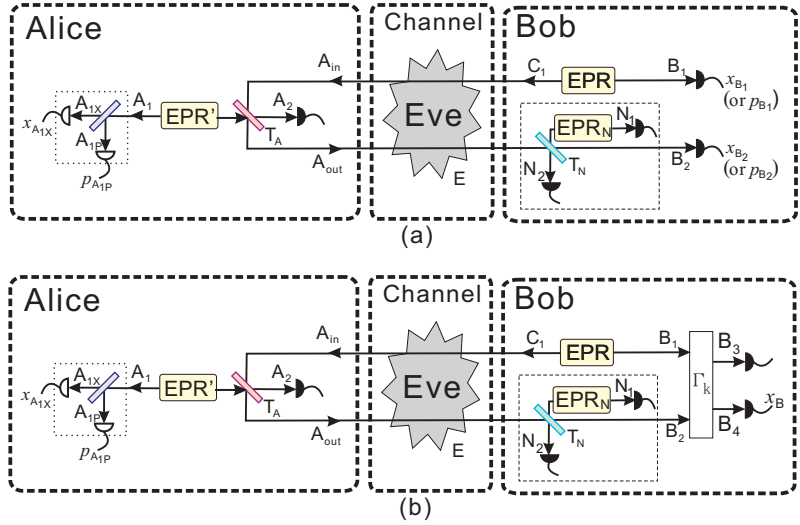}
\caption{\label{fig1}(a) The EB scheme of $\textrm{Hom}^2_{\textrm{N}}$ protocol. Bob keeps one half of the EPR pair (EPR) and sends the other half to Alice. Alice measures one mode of her EPR pair ($\rm{EPR}^\prime$) and one mode $A_2$ from a beam splitter $T_A$. The other mode from this beam splitter is returned back to Bob. The letters (e.g. $\textrm{B}_1$) beside arrows: the mode at the arrow; E: Eve's whole mode; the dashed box at $A_1$: the heterodyne detection; the dashed box at $B_2$: the added noise. (b) The equivalent scheme to \fref{fig1} (a) with postprocessing. Bob uses a symplectic transformation $\Gamma_k$ to change the modes $B_2$ and $B_1$ into $B_3$ and $B_4$.}
\end{figure}

The entanglement-based (EB) scheme of the two-way CV-QKD protocol $\textrm{Hom}_\textrm{N}^2$ with added detection noise is shown in \fref{fig1}(a), where the dashed box at $B_2$ is the added noise and the other part is our original  two-way protocol $\textrm{Hom}_\textrm{M}^2$ \cite{ourpaper}. The added noise is equivalent to an Einstein-Podolsky-Rosen (EPR) pair with the variance of $V_N$ coupled into the channel by a beam splitter with the transmittance of $T_N$. The protocol $\textrm{Hom}_\textrm{N}^2$ is described as follows.

\emph{Step one}. Bob initially keeps one mode $B_1$ of an EPR pair with the variance of $V$ while sending the other mode $C_1$ to Alice through the forward channel.

\emph{Step two}.  Alice measures one mode $A_1$ of her EPR pair (variance: $V_A$) to get the variables \{$x_{A_{1X}}$, $p_{A_{1P}}$\} with a heterodyne detection, and couples the other mode of her EPR pair  with the received mode $A_{in}$ from Bob by a beam splitter (transmittance: $T_A$). One output mode $A_2$ of the beam splitter is measured with homodyne detection and the other output mode $A_{out}$ is sent back to Bob through the backward channel.

\emph{Step three}. With a beam splitter (transmittance: $T_N$), Bob couples another EPR pair (variance: $V_N$) which is equivalent to the added noise with his received mode. The two modes $N_1$ and $N_2$ of this EPR pair are measured. Bob performs  homodyne detections on both modes $B_1$ and $B_2$ to get the variables $x_{B_1}$ (or $p_{B_1}$) and $x_{B_2}$ (or $p_{B_2}$), respectively.

\emph{Step four}. Alice and Bob implement the reconciliation and privacy amplification \cite{postprocessing,privacyamplificationCHBennett}. In this step, the measurement values of the modes $B_2$, $B_1$, $A_2$, $A_1$, $N_2$ and $N_1$ are used to estimate the channel's parameters and Bob uses $x_{B}=x_{B_2}-kx_{B_1}$ ($p_{B}=p_{B_2}+kp_{B_1}$) to construct the optimal estimation to Alice's corresponding variables $x_{A_{1X}}$ ($p_{A_{1P}}$), where $k$ is the channel's total transmittance.

The prepare-and-measure (PM) scheme of the two-way protocol can be equivalent to the EB scheme. In fact, Alice heterodyning one half of the EPR pair at $A_1$ is equivalent to remotely preparing a coherent state, and Bob performing homodyne detection on one half of the EPR pair at $B_1$ is equivalent to remotely preparing a squeezed state \cite{Grosshans03reventanglecloner}. The homodyne detection preceded by an EPR pair coupled by a beam splitter at $B_2$ is equivalent to Bob's real homodyne detection with efficiency $T_N$ and electronic noise \cite{8Cerfaddednoiseenhancekey}. Note that $x$ and $p$ quadratures are randomly measured in homodyne detection and only $x$ quadrature is analysed in the following.

\section{The analysis of the security against general collective attack}
First, we show that the Gaussian attack is optimal to the two-way protocol $\textrm{Hom}^2_{\textrm{N}}$ in general collective attack. In \fref{fig1}(a), since all modes of Alice and Bob are measured, Eve can get the purification of the state of Alice and Bob. In addition, the $x$ and $p$ quadratures of Alice and Bob's modes are not mixed via heterodyne or homodyne detection and Alice and Bob use the second-order moments of the quadratures to bound Eve's information. Therefore, the two-way protocol $\textrm{Hom}^2_{\textrm{N}}$ can satisfy the requirement of optimality of Gaussian collective attack (i.e., continuity, invariance under local Gaussian unitary and strong subadditivity) \cite{doctorRaul}. When the corresponding covariance matrix of the state $B_2B_1N_2N_1A_2A_1$ is known for Alice and Bob, the Gaussian attack is optimal \cite{14NJCerf2006,optimalMNavascues,optimalMMWolf,optimalLeverrier}. Therefore, only Eve's Gaussian collective attack is  needed to be considered in the following security analysis.

In \fref{fig1}(a), the secret key rate of the two-way protocol $\textrm{Hom}^2_{\textrm{N}}$ in reverse reconciliation is \cite{KRB1JLodewyck,KRB2ALeverrier,13Holevo,Grosshans03nature}
\begin{eqnarray}\label{KR}
K_R&=&\beta I(B:A)-I(B:E) \nonumber \\
&=&\frac{1}{2}\beta \log_2\frac{V_{A^M}}{V_{A^M|x_B}}-S(E)+S(E|x_B),
\end{eqnarray}
where $\beta$ is the reconciliation efficiency, $I(B:A)$ [$I(B:E)$] is the mutual information between Bob and Alice (Eve), $V_{A^M}$ and $V_{A^M|x_B}$ are Alice's variance and conditional variance, $S(E)$ and $S(E|x_B)$ are Eve's von Neumann entropy and conditional von Neumann entropy on Bob's data, respectively. In the following, $S(E)$ and $S(E|x_B)$ are calculated by the methods in \cite{QYCai}.

For Gaussian state, the entropy can be calculated from its corresponding covariance matrix \cite{twomodeeigenvalueJPBASerafini}. Since the state $B_2B_1N_2N_1A_2A_1E$ is a pure state, then $S(E)=S(B_2B_1N_2N_1A_2A_1)$. The corresponding covariance matrix of the state $B_2B_1N_2N_1A_2A_1$ is
\begin{eqnarray}\label{rhoAB}
\Gamma_{B_2B_1N_2N_1A_2A_1}=\left(
\begin{array}{cccccc}
 \gamma_{B_2} & C_1 & C_2  & C_3  & C_4  & C_5  \\
 C_1 &  V\mathbb{I}  & C_6 & 0    & C_7  & 0  \\
 C_2 & C_6 & \gamma_{N_2}  & C_8  & C_9  & C_{10} \\
 C_3 & 0 & C_8  & V_N\mathbb{I}   & 0    & 0    \\
 C_4 & C_7  & C_9 & 0 & \gamma_{A_2}  & C_{11} \\
 C_5 & 0 & C_{10}  &   0 & C_{11} & V_A\mathbb{I}
 \end{array}
\right),
\end{eqnarray}
where $\mathbb{I}$ is a $2\times2$ identity matrix, the diagonal elements correspond to the variances of $x$ and $p$ quadratures of the modes $B_2$, $B_1$, $N_2$, $N_1$, $A_2$ and $A_1$ in turn, e.g. $\gamma_{B_2}=\textrm{diag}(\langle x_{B_2}^2\rangle, \langle p_{B_2}^2\rangle)$, and the nondiagonal elements correspond to the covariances between modes, e.g. $C_1=\textrm{diag}(\langle x_{B_2}x_{B_1}\rangle,\langle p_{B_2}p_{B_1}\rangle)$. Therefore, Eve's entropy \cite{HolevoSi}
\begin{equation}\label{SE}
S(E)=\sum^6_{i=1} G(\lambda _i)=\sum^6_{i=1} G\left(f_{\lambda _i}(\alpha_{mn})\right),
\end{equation}
where $
G(\lambda _i)=\frac{\lambda _i+1}{2}\log\frac{\lambda _i+1}{2}-\frac{\lambda _i-1}{2}\log\frac{\lambda _i-1}{2},
$ 
and $\lambda_i=f_{\lambda _i}(\alpha_{mn})$ is the symplectic eigenvalue of $\Gamma\!_{B_2B_1N_2N_1A_2A_1}$ which is the function of the element $\alpha_{mn}$ of $\Gamma\!_{B_2B_1N_2N_1A_2A_1}$ \cite{eigenvalue,eigenvalueSPirandola}, seen in \ref{appendixA}.

Bob uses $x_B=x_{B_2}-kx_{B_1}$ to estimate Alice's variable, which is equivalent to that Bob uses a symplectic transformation $\Gamma_k$ to change the modes $B_2$ and $B_1$ into the modes $B_4$ and $B_3$ where the $x$ quadrature of the mode $B_4$ is $x_{B_4}=x_B=x_{B_2}-kx_{B_1}$ \cite{ourpaper}, as shown in \fref{fig1}(b). Since \fref{fig1}(b) is equivalent to \fref{fig1}(a) with postprocessing, we use \fref{fig1}(b) to calculate $S(E|x_B)$ in the following.

After the symplectic transformation $\Gamma_k$, the corresponding covariance matrix of the mode $B_4B_3N_2N_1A_2A_1$ is
\begin{equation}
  \Gamma_{B_4B_3N_2N_1A_2A_1}=[\Gamma_k\oplus\mathbb{I}_4]\Gamma_{B_2B_1N_2N_1A_2A_1}[\Gamma_k\oplus\mathbb{I}_4]^T,
\end{equation}
where $\mathbb{I}_4\!\!=\!\oplus^4_{1}\mathbb{I}$, $\Gamma\!_k$ is a continuous-variable C-NOT gate \cite{doctorRaul,QND,QuantumComputationandQuantumInformationNielsen}
\begin{equation}
 \Gamma_k=\left(
\begin{array}{cccc}
1&0&-k&0\\
0&1&0&0\\
0&0&1&0\\
0&k&0&1
\end{array}
\right).
\end{equation}
Since the state $B_3N_2N_1A_2A_1E$ is a pure state when Bob gets $x_B$ by measuring the modes $B_4$, then $S(E|x_B)=S(B_3N_2N_1A_2A_1|x_B)$. The corresponding covariance matrix of the state $B_3N_2N_1A_2A_1$ conditioned on $x_B$ is \cite{doctorRaul,JEisertconditionalmatrix}
\begin{equation} \label{GammConditionxB}
  \Gamma_{B_3N_2N_1A_2A_1}^{x_B}=\gamma_{B_3N_2N_1A_2A_1}-C_{B_4}[X_x\gamma_{B_4}X_x]^{MP}C_{B_4},
\end{equation}
where $\gamma_{B_3N_2N_1A_2A_1}$ and $\gamma_{B_4}$ are the corresponding reduced matrixes of the states $B_3N_2N_1A_2A_1$ and $B_4$ in $\Gamma_{B_4B_3N_2N_1A_2A_1}$, respectively, $C_{B_4}$ is their correlation matrix, $X_x=diag(1,0)$ and $MP$ denotes the inverse on the range. Therefore, we have
\begin{equation}\label{S(E|xBpB)}
S(E|x_B)=\sum^5_{i=1} G(\lambda _i^\prime)=\sum^5_{i=1} G\left(f_{\lambda _i^\prime}(\alpha_{mn}^\prime)\right),
\end{equation}
where $\lambda_i^\prime=f_{\lambda _i^\prime}(\alpha_{mn}^\prime)$ is the symplectic eigenvalue of $\Gamma_{\!\!B_3N_2N_1A_2A_1}^{x_B}$ which is the function of the element $\alpha_{mn}^\prime$ of $\Gamma_{\!\!B_3N_2N_1A_2A_1}^{x_B}$ \cite{eigenvalue,eigenvalueSPirandola}, seen in \ref{appendixA}.

By substituting equations (\ref{SE}) and (\ref{S(E|xBpB)}) into equation (\ref{KR}), the secret key rate is obtained
\begin{equation} \label{KR(a{mn})}
K_R\!=\!\frac{1}{2}\beta \log_2\frac{V_{A^M}}{V_{A^M|x_B}}\!-\!\sum^6_{i=1}\! G\left(f_{\lambda _i}(\alpha_{mn})\right)\!+\!\sum^5_{i=1}\! G\left(f_{\lambda _i^\prime}(\alpha_{mn}^\prime)\right).
\end{equation}
In experiment, Alice and Bob can calculate the element $\alpha_{mn}$ and $\alpha_{mn}^\prime$ of equations (\ref{rhoAB}) and (\ref{GammConditionxB}) by the measurement values of the modes $B_2$, $B_1$, $N_2$, $N_1$, $A_2$ and $A_1$. Therefore, according to equation (\ref{KR(a{mn})}), the secret key rate in general collective attack is obtained without the assumption that the two channels are uncorrelated. The analytic representations of equation (\ref{KR(a{mn})}) is too complex to give here. We give a numerical simulation in the following.

\begin{figure}
\centering
\includegraphics[width=10.5cm]{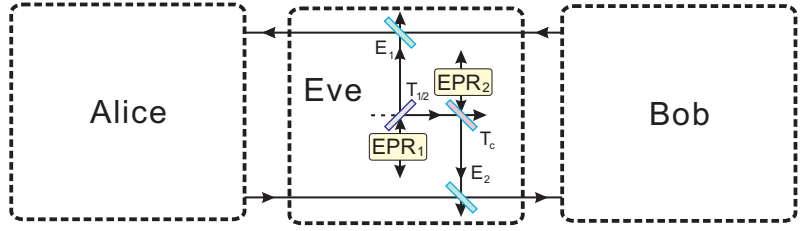}
\caption{\label{fig2correlatedchannels}The EB scheme of $\textrm{Hom}^2_{\textrm{N}}$ protocol against entangling cloner attacks on correlated channels. $E_1$, $E_2$: the modes introduced into the channels; $T_{1/2}$: half beam splitter; $T_c$: beam splitter. Alice and Bob are the same as \fref{fig1}(a).}
\end{figure}

\begin{figure}
\centering
\includegraphics[width=10.5cm]{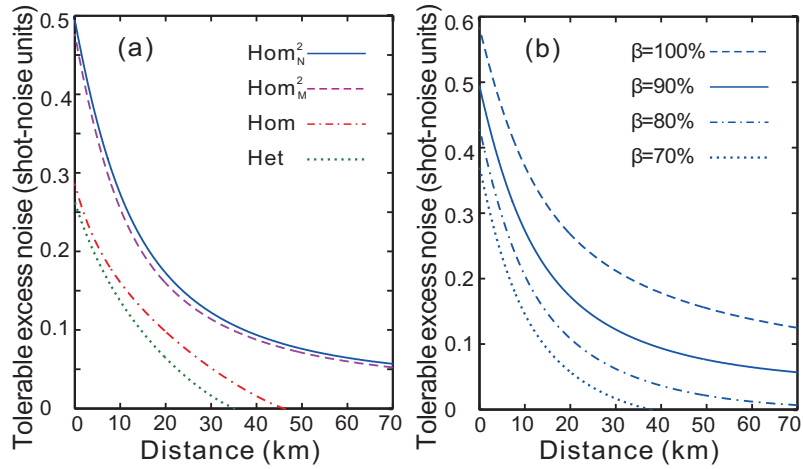}
\caption{\label{fig4Tolerablenoiseuncorrelated}(a) Tolerable excess noise as a function of the transmission distance for $\textrm{Hom}_\textrm{N}^2$, $\textrm{Hom}_\textrm{M}^2$, $\textrm{Hom}$ and $\textrm{Het}$ protocols, where $\beta=90\%$. (b) Tolerable excess noise as a function of the transmission distance for $\textrm{Hom}_\textrm{N}^2$ protocol, where $\beta=100\%, 90\%, 80\%, 70\%$. The curves of (a) and (b) are plotted for $n_c=0$, $T_A=0.8$ and $V_A=V=20$.}
\end{figure}

\begin{figure}
\centering
\includegraphics[width=10.5cm]{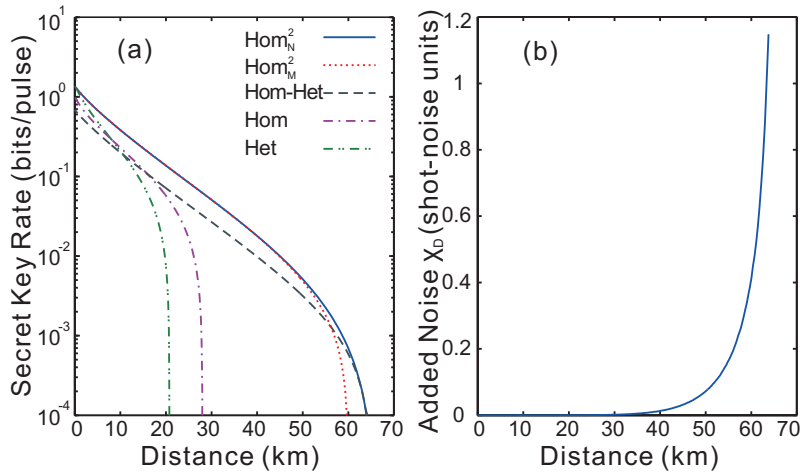}
\caption{\label{fig5kuncorrelated}(a) Secret key rate as a function of the transmission distance for $\textrm{Hom}_\textrm{N}^2$, $\textrm{Hom}_\textrm{M}^2$, $\textrm{Hom-Het}$, $\textrm{Hom}$ and $\textrm{Het}$ protocols, where $n_c=0$, $\varepsilon=0.06$, $\beta=90\%$, $T_A=0.8$ and $V_A=V=20$. (b) Optimal choice of the added noise $x_D$.}
\end{figure}

\begin{figure}
\centering
\includegraphics[width=10.5cm]{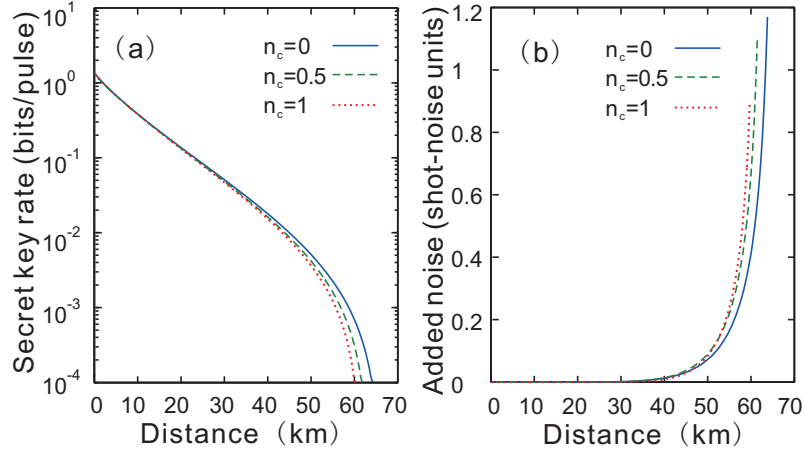}
\caption{\label{fig3kcorrelated}(a) Secret key rate as a function of the transmission distance for $\textrm{Hom}_\textrm{N}^2$ protocol, where $\varepsilon=0.06$, $\beta=90\%$, $T_A=0.8$, $V_A=V=20$ and $n_c=0,0.5,1$. (b) Optimal choice of the added noise $x_D$.}
\end{figure}

\begin{figure}
\centering
\includegraphics[width=10.5cm]{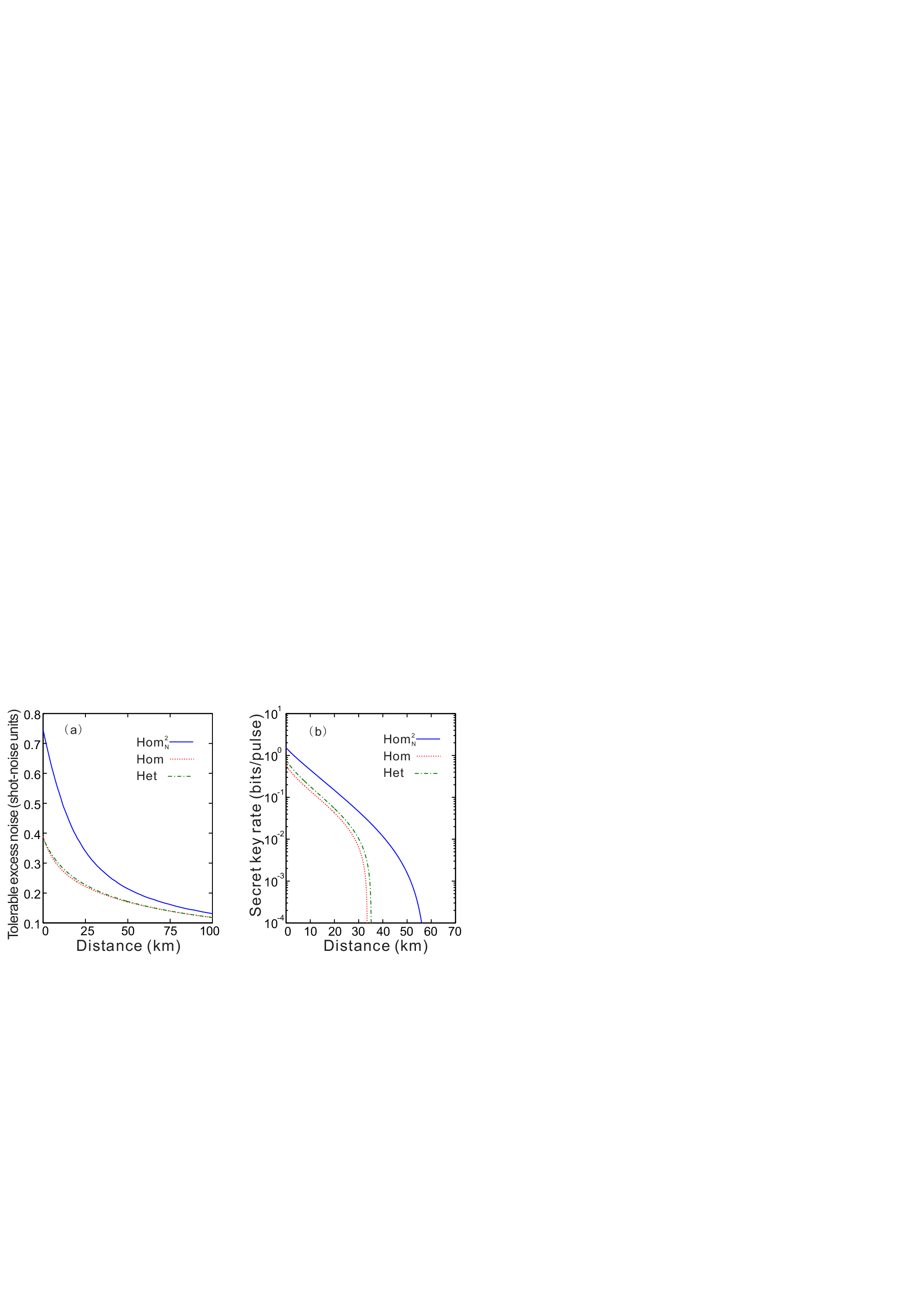}
\caption{\label{fig6ek}(a) Tolerable excess noise as a function of the transmission distance for high modulation for $\textrm{Hom}_\textrm{N}^2$ protocol. (b) Secret key rate as a function of the transmission distance for high modulation for $\textrm{Hom}_\textrm{N}^2$ protocol, where $\varepsilon=0.2$. The curves of (a) and (b) are plotted for $V_A=V=1000$, $n_c=0$, $T_A=0.8$ and $\beta=1$.}
\end{figure}

\section{Numerical simulation and discussion of collective entangling cloner attacks on correlated and uncorrelated channels}
For simplicity in numerical simulation, when there is no Eve, the forward and the backward channels are assumed to be independent with the identical transmittances $T$ and noises referred to the input $\chi=\varepsilon+(1-T)/T$, where  $\varepsilon$ is the channel excess noises referred to the input. It is equivalent to Eve implementing two independent collective entangling cloner attacks which are a Gaussian collective attack investigated in detail in \cite{channelSPirandola,channelASHolevo}. When Eve implements more complicated two-mode attack \cite{6Pirandola}, the correlation between the two channels is induced. \Fref{fig2correlatedchannels} shows that Eve implements two correlated entangling cloner attacks. On condition that Eve introduces the equivalent variances of the modes $E_1$ and $E_2$ into the two channels, the noise referred to the input of the backward channel is $\chi_2=\chi+2n_c\sqrt{T_AT}\varepsilon$, where the second item on the right-hand side is induced extra by the correlation between the two channels, i.e., the part of the mode introduced into the backward channel correlating with the forward channel interferes with the mode from Alice, $n_c=\sqrt{1-T_c}$ is the coefficient representing the degree of the correlation, e.g., $n_c=0$ represents that the two channels are uncorrelated. The added noise is $\chi_D=(1-T_N)V_N/T_N$. We can calculate the elements of equation (\ref{rhoAB})
\begin{eqnarray} \label{elementGmmaAB}
&\gamma_{B_2}=\{V\!_N - T\!_N V\!_N + T T\!_N [V\!_A - T\!_A V\!_A + T T\!_A (V + \chi) + \chi_2] \}  \mathbb{I},\nonumber \\
&\gamma_{N_2}=\{T\!_N V\!_N + T (1 - T\!_N) [V\!_A - T\!_A V\!_A + T T\!_A (V + \chi) + \chi_2] \}  \mathbb{I},\nonumber  \\
&\gamma_{A_2}= [T\!_A V\!_A + T (1- T\!_A) (V + \chi)]  \mathbb{I},\nonumber  \\
&C_1\!= -\eta C_6=T\sqrt{T\!_A T\!_N (V^2-1)} \sigma_{\!z},  \nonumber \\
&C_2\!=\!\sqrt{(1\!-\! T\!_N) T\!_N} \{V\!_N - T [V\!_A - T\!_A V\!_A + T T\!_A (V + \chi) + \chi_2]\}  \mathbb{I}, \nonumber \\
&C_3\!=\frac{1}{\eta}C_8=\sqrt{(1- T\!_N) (V_N^2-1)}  \sigma_{\!z}, \nonumber \\
&C_4\!=-\eta C_9=\!\!\!\sqrt{T (\!1\!-\! T\!_A\!)  T\!_N} \left[\!\!\sqrt{T\!_A}V\!_A\! -\! T\! \sqrt{T\!_A}(\!V \!+\! \chi) \!-\!n_c\!\sqrt{T}\varepsilon\right]  \mathbb{I}, \nonumber \\
&C_5\!=-\eta C_{10}=\sqrt{T (1- T\!_A) T\!_N (V^2\!_A-1)} \sigma_{\!z}, \nonumber \\
&C_7\!=-\sqrt{T (1 - T\!_A) (V^2-1)} \sigma_{\!z}, \nonumber \\
&C_{11}\!=\sqrt{T\!_A (V^2\!_A-1)} \sigma_{\!z},
\end{eqnarray}
and
\begin{equation}\label{IBA}
I(\!B\!:\!A\!)\!=\!\frac{1}{2}\!\log_2 \!\frac{T^2 T_A F + T (V_A - T_A V_A + \chi_2) + \chi_D}{T (1 - T_A + \chi_2) + T^2 T_A F + \chi_D},
\end{equation}
where
$
\sigma\!_z\!=diag(1,-1),\eta\!=\!\sqrt{T\!_N/(1\!-\!T\!\!_N)}, F\!=\!\!2V \!-\! 2\!\sqrt{V^2\!-\!1}\!+ \! \chi.
$
The typical fiber channel loss is assumed to be 0.2 dB/km. $V$ and $\varepsilon$ are in shot-noise units. Substituting equations (\ref{elementGmmaAB}) and (\ref{IBA}) into equation (\ref{KR(a{mn})}), the optimal secret key rate $K_R$ and the optimal tolerable excess noise $\varepsilon$ of the two-way protocol $\textrm{Hom}^2_\textrm{N}$ can be obtained by adjusting the added noise $\chi_D$.

When $n_c=0$, the two channels are uncorrelated, which is equivalent to Eve implementing two independent Gaussian cloner attacks. For comparison, the heterodyne protocol ($\textrm{Het}$) \cite{het} and the homodyne protocol ($\textrm{Hom}$) \cite{Grosshans03nature} of one-way CV-QKD protocol with coherent state and the original modified two-way protocols $\textrm{Hom}^2_\textrm{M}$ and $\textrm{Hom-Het}_\textrm{M}$ \cite{ourpaper} are also given in figures \ref{fig4Tolerablenoiseuncorrelated}(a) and \ref{fig5kuncorrelated}(a). \Fref{fig4Tolerablenoiseuncorrelated}(a) shows the tolerable excess noise as a function of the transmission distance, where $V_A=V=20$, $T_A=0.8$, $\beta=90\%$ and $n_c=0$. The proper added noise $\chi_D$ is chosen to make $\varepsilon$ of $\textrm{Hom}^2_\textrm{N}$ protocol optimal. The numerical simulation result indicates that the tolerable excess noise of the two-way protocol with added noise is more than that without added noise and surpasses that of the one-way CV-QKD protocol. Therefore, it indicates that properly added noise is useful to enhance $\varepsilon$ in the two-way protocol. \Fref{fig4Tolerablenoiseuncorrelated}(b) shows the tolerable excess noise $\varepsilon$ of $\textrm{Hom}^2_\textrm{N}$ protocol with different $\beta$, which indicates that the tolerable excess noise $\varepsilon$ increases with the increase of $\beta$.

\Fref{fig5kuncorrelated}(a) shows the secret key rate as a function of the transmission distance, where $V=V_A=20$, $T_A=0.8$, $\varepsilon=0.06$, $\beta=90\%$ and $n_c=0$. To make the secret key rate of $\textrm{Hom}_\textrm{N}^2$ optimal, the proper added noise $\chi_D$ is chosen, as shown in \fref{fig5kuncorrelated}(b). In \fref{fig5kuncorrelated}(a), the simulation result indicates that the two-way protocol with added noise has higher secret key rate than that without added noise. Especially, the achievable transmission distance of the two-way protocol $\textrm{Hom}^2_\textrm{N}$ is over 60 km when $\beta$ is $90\%$, which is much longer than that of the one-way protocol. The reason is that the added noise not only lowers the mutual information between Alice and Bob, but also lowers that between Bob and Eve. When the effect on Eve is more than that on Alice and Bob, the secret key rate is enhanced.

When $n_c \neq 0$, the two channels are correlated. \Fref{fig3kcorrelated}(a) shows the secret key rate as a function of the transmission distance for $\textrm{Hom}_\textrm{N}^2$ protocol with different $n_c$. Considering the practical experiment \cite{detectnoiseJLodewyck,KRB1JLodewyck,TSymulexperiment}, we choose $\varepsilon=0.06$, $\beta=90\%$, $T_A=0.8$, $V_A=V=20$ and $n_c=0,0.5,1$. To make $K_R$ optimal, the proper added noise $\chi_D$ is chosen, as shown in \fref{fig3kcorrelated}(b). In \fref{fig3kcorrelated}(a), the simulation result indicates that the  distance of the secret key distribution decreases with the increase of $n_c$. The reason is that the correlation between the two channels induces the change of the excess noise in the backward channel, which affects the secret key rate. \Fref{fig3kcorrelated}(a) shows that the decrease of the secret key rate induced by this effect is small. In addition, comparing with  the one-way protocol in \fref{fig5kuncorrelated}(a), despite the transmission distance of the two-way protocol decreases slightly due to the correlation, the performance of the two-way protocol is still far beyond that of the one-way protocols. \Fref{fig3kcorrelated}(b) shows that the optimal added noise decreases with the decrease of $n_c$.

In the following, we compare the two-way protocol with the one-way protocols in high modulation. Figures \ref{fig6ek}(a) and (b) show the tolerable excess noise and the secret key rate as a function of the transmission distance for high modulation, where $V_A=V=1000$, $T_A=0.8$, $\beta=1$ and $n_c=0$. The proper added noise $\chi_D$ is chosen to make the tolerable excess noise and the secret key rate of $\textrm{Hom}_\textrm{N}^2$ protocol optimal. The numerical simulation result indicates that both the tolerable excess noise and the secret key rate of the two-way protocol with added noise are much more than that of the one-way CV-QKD protocols for high modulation.

\section{Conclusion}
In conclusion, we improve the two-way CV-QKD protocol by adding a proper noise on Bob's detection side. The security of the two-way CV-QKD protocol with added noise in homodyne detection against general collective attack is analysed. The numerical simulation under the collective entangling cloner attack is given for the correlated and the uncorrelated channels. The simulation result indicates that despite the secret key rate for the correlated channels is slightly lower than that for the uncorrelated channels when Eve inputs equivalent variance of the modes into the two channels, the performance of the two-way protocol is still far beyond that of the one-way protocols. In addition, the properly added noise is beneficial for enhancing the secret key rate and the tolerable excess noise of the two-way CV QKD. The optimal tolerable excess noise of the two-way CV QKD with added noise is much more than that of the one-way CV QKD. With the reasonable reconciliation efficiency of $90\%$, the two-way CV QKD with added noise allows the distribution of secret keys over 60 km fibre distance, which is difficult to reach for the one-way CV-QKD protocols with Gaussian modulation in experiment.

\section*{Acknowledgments}
This work is supported by the National Science Fund for Distinguished Young Scholars of China (Grant No. 61225003), National Natural Science Foundation of China (Grant No. 61101081), and the National Hi-Tech Research and Development (863) Program.

\appendix
\section{\label{appendixA}The calculation of eigenvalues}

\setcounter{equation}{0}

\renewcommand{\theequation}{A.\arabic{equation}}

We use $\alpha^{\prime\prime}_{mn}$ to denote the elements of the corresponding covariance matrix $\Gamma_n$ of a $n$-mode state.
The symplectic invariants $\{\vartriangle_{n,j}\}$ of $\Gamma_n$ for $j=1,..., n$ are defined as \cite{eigenvalue}
\begin{equation}
  \vartriangle_{n,j}=M_{2j}(\Omega\Gamma_n),
\end{equation}
where $\Omega=\oplus_1^ni\sigma_y$ ($\sigma_y$ standing for the $y$ Pauli matrix) and $M_{2j}(\Omega\Gamma)$ is the principal minor of order $2j$ of the $2n\times2n$ matrix $\Omega\Gamma$ which is the sum of the determinants of all the $2j\times2j$  submatrices of $\Omega\Gamma_n$ \cite{eigenvalue,eigenvalueSPirandola}.

The symplectic eigenvalues of the matrix corresponding to a four-mode state are the solution of the four-order equation on the symplectic invariants \cite{ourpaper,doctorRaul,eigenvalue}
\begin{eqnarray} \label{lamda1234}
  f^2\!\!\!\!_{\lambda_{1,2}^{\prime\prime}}\!(\!\alpha^{\prime\prime}_{mn}\!)\!\!=\!\!\frac{\vartriangle_{4\!,1}}{4}\!-\!\frac{1}{2}\sqrt{\zeta\!+\!\Theta}\pm\frac{1}{2}\!\sqrt{\!2\zeta\!-\!\Theta\!-\!\frac{\vartriangle_{4\!,1}^3 \!- 4 \!\!\vartriangle_{4\!,1} \vartriangle_{4\!,2} \!+\! 8 \!\!\vartriangle_{4\!,3}}{4\sqrt{\zeta\!+\!\Theta}}} , \nonumber\\
   f^2\!\!\!\!_{\lambda_{3,4}^{\prime\prime}}\!(\!\alpha^{\prime\prime}_{mn}\!)\!\!=\!\!\frac{\vartriangle_{4\!,1}}{4}\!+\!\frac{1}{2}\sqrt{\zeta\!+\!\Theta}\pm\frac{1}{2}\!\sqrt{\!2\zeta\!-\!\Theta\!+\!\frac{\vartriangle_{4\!,1}^3 \!- 4 \!\!\vartriangle_{4\!,1} \vartriangle_{4\!,2} \!+\! 8\!\!\vartriangle_{4\!,3}}{4\sqrt{\zeta\!+\!\Theta}}},\nonumber\\
\end{eqnarray}
where
\begin{eqnarray}
 \!\!\!\! &&\zeta=\frac{\vartriangle_{4\!,1}^2}{4}\!-\!\frac{2\!\!\vartriangle_{4\!,2}}{3},  \qquad \Theta=\frac{2^{\frac{1}{3}}H}{3J}+\frac{J}{3\cdot 2^{\frac{1}{3}}},  \nonumber\\
\!\!\!\!  &&H=\vartriangle_{4,2}^2\! - 3\!\! \vartriangle_{4,1} \vartriangle_{4,3}\! + 12 \!\! \vartriangle_{4,4},
  \qquad J = \left(L+\sqrt{L^2-4 H^3}\right)^{\frac{1}{3}},  \nonumber\\
  \!\!\!\!  &&L\!\!=\!\!2\!\! \vartriangle_{4,2}^3\! \!- 9\!\! \vartriangle_{4,1} \vartriangle_{4,2} \vartriangle_{4,3}\!\! + 27\!\! \vartriangle_{4,3}^2 \!\!+ 27\! \! \vartriangle_{4,1}^2 \vartriangle_{4,4}\!\! - 72\!\! \vartriangle_{4,2} \vartriangle_{4,4}.
\end{eqnarray}

From equation (\ref{rhoAB}), the covariance matrix $\Gamma_{N_1N_2B_2B_1A_2A_1}$ of the modes $N_1N_2B_2B_1A_2A_1$ can be obtained by permuting the corresponding elements of $\Gamma_{B_2B_1N_2N_1A_2A_1}$. Applying a unitary transformation $S=\mathbb{I}\oplus\Gamma_{T_N}\oplus\mathbb{I}\oplus\mathbb{I}\oplus\mathbb{I}$ to equation (\ref{rhoAB}), we can obtain
\begin{equation}
  S^T\Gamma_{N_1N_2B_2B_1A_2A_1}S=\left(
\begin{array}{cc}
\Gamma_{N_1N^\prime_2}&0\\
0&\Gamma_{B^\prime_2B_1A_2A_1}
\end{array}
\right),
\end{equation}
where
\begin{eqnarray}
&&\Gamma_{T_N}=\!\left(\!\!
\begin{array}{cc}
\sqrt{T\!_N}\mathbb{I}  & \sqrt{1\!-\!T\!_N}\mathbb{I}\\
-\sqrt{1\!-\!T\!_N}\mathbb{I} & \sqrt{T\!_N}\mathbb{I}
\end{array}
\!\!\right),\qquad
\Gamma\!\!_{N_1N^\prime_2}\!\!=\!\left(\!\!\!\!
\begin{array}{cc}
V_N\mathbb{I}       \!\!\!\!\!\!& \sqrt{V_{\!\!N}^2\!-\!1}\sigma_z\\
\sqrt{V_{\!\!N}^2\!-\!1}\sigma_z   \!\!\!\!\!\!&    V_N\mathbb{I}
\end{array}
\!\!\right),\nonumber\\
&&\Gamma\!\!_{B^\prime_2B_1A_2A_1}\!\!=\!\left(\!\!\begin{array}{cccc}
 \gamma^\prime_{B_2} & C^\prime_1 &  C^\prime_4  & C^\prime_5  \\
 C^\prime_1 &  V\mathbb{I} & C_7   & 0  \\
 C^\prime_4 & C_7  &  \gamma_{A_2} & C_{11} \\
 C^\prime_5 & 0    & C_{11} & V_A\mathbb{I}
 \end{array}
\!\!\right),
\end{eqnarray}
and $\gamma^\prime_{B_2}=[\gamma_{B_2}-(1-T_N)V_N]/T_N$, $C^\prime_i=C_i/\sqrt{T_N}$ for $i=1, 4, 5$. Therefore, the eigenvalues of $\Gamma_{B_2B_1N_2N_1A_2A_1}$ are $\lambda_i=f_{\lambda_{1,2,3,4}}(\alpha_{mn}),1,1$, where $f_{\lambda_{1,2,3,4}}(\alpha_{mn})$ are the eigenvalues of $\Gamma_{B^\prime_2B_1A_2A_1}$ given by equation (\ref{lamda1234}).

The symplectic invariants of $\Gamma_{B_3N_2N_1A_2A_1}^{x_B}$ are denoted as $\vartriangle_{5\!,j}$ for j=1...5. It can be proved that $1\!-\!\!\vartriangle_{5\!,1}\!\!\!+\!\vartriangle_{5\!,2}\!\!-\!\!\vartriangle_{5\!,3}\!\!+\!\!\vartriangle_{5\!,4}\!\!-\!\!\vartriangle_{5\!,5}\!=\!0$.
Therefore, one of the eigenvalues of $\Gamma_{B_3N_2N_1A_2A_1}^{x_B}$ is 1 and the others have the same forms of equation (\ref{lamda1234}), which needs the replacement
$
\vartriangle_{4\!,1}=\vartriangle_{5\!,1}\!\!-1,
\vartriangle_{4\!,2}=\vartriangle_{5\!,2}\!\!-\vartriangle_{5\!,1}\!\!+1,
\vartriangle_{4\!,3}=\vartriangle_{5\!,4}\!\!-\vartriangle_{5\!,5},
\vartriangle_{4\!,4}=\vartriangle_{5\!,5}.
$


\section*{References}
\begin{thebibliography}{99}

\bibitem{Scarani} Scarani V, Bechmann-Pasquinucci H, Cerf N J, Du\v{s}ek M, L\"{u}tkenhaus N and Peev M 2009 {\it Rev. Mod. Phys.} {\bf 81} 1301

\bibitem{BB84} Bennett C H and Brassard G 1984 Quantum cryptography: public key distribution and coin tossing {\it Proc. IEEE Int. Conf. on Computers, Systems and Signal Proceedings} (New York: IEEE Press) pp~175--9

\bibitem{postselectionSilberhorn} Silberhorn C, Ralph T C, L\"{u}tkenhaus N and Leuchs G 2002 {\it Phys. Rev. Lett.} {\bf 89} 167901

\bibitem{improvingdistanceALeverrier} Blandino R, Leverrier A, Barbieri M, Etesse J, Grangier P and Tualle-Brouri R 2012 {\it Phys. Rev. A} {\bf 86} 012327

\bibitem{squeezedstateprotocolMHillery} Hillery M 2000 {\it Phys. Rev. A} {\bf 61} 022309

\bibitem{2Grosshans02} Grosshans F and Grangier P 2002 {\it Phys. Rev. Lett.} {\bf 88} 057902



\bibitem{6Pirandola} Pirandola S, Mancini S, Lloyd S and Braunstein S L 2008 {\it Nature Phys.} {\bf 4} 726

\bibitem{8Pirandola} Weedbrook C, Pirandola S, Garc\'{\i}a-Patr\'{o}n R, Cerf N J, Ralph T C, Shapiro J H and Lloyd S 2012 {\it Rev. Mod. Phys.} {\bf 84} 621

\bibitem{ourpaper} Sun M, Peng X, Shen Y and Guo H 2012 {\it Int. J. Quant. Inf.} {\bf 10} 1250059







\bibitem{sourcenoiseRFilip} Filip R 2008 {\it Phys. Rev. A} {\bf 77} 022310

\bibitem{sourcenoiseRFilip2010} Usenko V C and Filip R 2010 {\it Phys. Rev. A} {\bf 81} 022318

\bibitem{sourcenoiseaddednoiseenhancekeyWeedbrook} Weedbrook C, Pirandola S, Lloyd S and Ralph T C 2010 {\it Phys. Rev. Lett.} {\bf 105} 110501
\bibitem{sourcenoiseYongShen} Shen Y, Yang J and Guo H 2009 {\it J. Phys. B: At. Mol. Opt. Phys.} {\bf 42} 235506
\bibitem{sourcenoiseYShen} Shen Y, Peng X, Yang J and Guo H 2011 {\it Phys. Rev. A} {\bf 83} 052304


\bibitem{detectnoiseJLodewyck} Lodewyck J, Debuisschert T, Tualle-Brouri R and Grangier P 2005 {\it Phys. Rev. A} {\bf 72} 050303

\bibitem{KRB1JLodewyck} Lodewyck J, Bloch M, Garc\'{\i}a-Patr\'{o}n R, Fossier S, Karpov E, Diamanti E, Debuisschert T, Cerf N J, Tualle-Brouri R, McLaughlin S W and Grangier P 2007 {\it Phys. Rev. A} {\bf 76} 042305




\bibitem{8Cerfaddednoiseenhancekey} Garc\'{\i}a-Patr\'{o}n R and Cerf N J 2009 {\it Phys. Rev. Lett.} {\bf 102} 130501

\bibitem{Weedbrookaddednoiseenhancekey} Weedbrook C, Pirandola S and Ralph T C 2012 {\it Phys. Rev. A} {\bf 86} 022318

\bibitem{Pirandolaaddednoiseenhancekey} Pirandola S, Garc\'{\i}a-Patr\'{o}n R, Braunstein S L and Lloyd S 2009 {\it Phys. Rev. Lett.} {\bf 102} 050503


\bibitem{Renneraddednoiseenhancekey} Renner R, Gisin N and Kraus B 2005 {\it Phys. Rev. A} {\bf 72} 012332

\bibitem{Renesaddednoiseenhancekey} Renes J M and Smith G 2007 {\it Phys. Rev. Lett.} {\bf 98} 020502



\bibitem{MWangandWPan} Wang M and Pan W 2010 {\it Phys. lett. A} {\bf 374} 2434



\bibitem{correlatednoiseZengGuihua2012} Huang P, Zhu J, He G  and Zeng G 2012 {\it J. Phys. B: At. Mol. Opt. Phys.} {\bf 45} 135501


\bibitem{correlatednoiseZengGuihua2011} Huang P, He G, Lu Y and Zeng G 2011 {\it Phys. Scr.} {\bf83} 015005


\bibitem{Grosshans03reventanglecloner} Grosshans F, Cerf N J, Wenger J, Tualle-Brouri R and Grangier P 2003 {\it Quantum Inf. Comput.} {\bf 3} 535

\bibitem{CWeedbrookentanglecloner} Weedbrook C, Grosse N B, Symul T, Lam P K and Ralph T C 2008 {\it Phys. Rev. A} {\bf 77} 052313





\bibitem{postprocessing} Assche G V 2006 {\it Quantum Cryptography and Secret-Key Distillation} (Cambridge: Cambridge University Press)

\bibitem{privacyamplificationCHBennett} Bennett C H, Brassard G, Cr\'{e}peau C and Maurer U M 1995 {\it IEEE Trans. Inf. Theory} {\bf 41} 1915




\bibitem{doctorRaul} Garc\'{\i}a-Patr\'{o}n R 2007 Ph.D. thesis, Universit\'{e} Libre de Bruxelles


\bibitem{14NJCerf2006} Garc\'{\i}a-Patr\'{o}n R and Cerf N J 2006 {\it Phys. Rev. Lett.} {\bf 97} 190503

\bibitem{optimalMNavascues} Navascu\'{e}s M, Grosshans F and Ac\'{\i}n A 2006 {\it Phys. Rev. Lett.} {\bf 97} 190502

\bibitem{optimalMMWolf} Wolf M M, Giedke G and Cirac J I 2006 {\it Phys. Rev. Lett.} {\bf 96} 080502

\bibitem{optimalLeverrier} Leverrier A and Grangier P 2010 {\it Phys. Rev. A} {\bf 81} 062314




\bibitem{KRB2ALeverrier} Leverrier A, All\'{e}aume R, Boutros J, Z\'{e}mor G and Grangier P 2008 {\it Phys. Rev. A} {\bf 77} 042325

\bibitem{13Holevo} Holevo A S 1973 {\it Probl. Inf. Transm.} {\bf 9} 177

\bibitem{Grosshans03nature} Grosshans F, Assche G V, Wenger J, Brouri R, Cerf N J and Grangier P 2003 {\it Nature} {\bf 421} 238






\bibitem{QYCai} Lu H, Fung C F, Ma X and Cai Q 2011 {\it Phys. Rev. A} {\bf 84} 042344



\bibitem{twomodeeigenvalueJPBASerafini} Serafini A, Illuminati F and De Siena S 2004 {\it J. Phys. B: At. Mol. Opt. Phys.} {\bf 37} L21



\bibitem{HolevoSi} Holevo A S, Sohma M and Hirota O 1999 {\it Phys. Rev. A} {\bf 59} 1820


\bibitem{eigenvalue} Serafini A 2006 {\it Phys. Rev. Lett.} {\bf 96} 110402


\bibitem{eigenvalueSPirandola} Pirandola S, Serafini A and Lloyd S 2009 {\it Phys. Rev. A} {\bf 79} 052327




\bibitem{QND} Yoshikawa J I, Miwa Y, Huck A, Andersen U L, van Loock P and Furusawa A 2008 {\it Phys. Rev. Lett.} {\bf 101} 250501


\bibitem{QuantumComputationandQuantumInformationNielsen} Nielsen M A and Chuang I L 2000 {\it Quantum Computation and Quantum Information} (Cambridge: Cambridge University Press)





\bibitem{JEisertconditionalmatrix} Eisert J and Plenio M B 2003 {\it Int. J. Quant. Inf.} {\bf 1} 479







\bibitem{channelSPirandola} Pirandola S, Braunstein S L and Lloyd S 2008 {\it Phys. Rev. Lett.} {\bf 101} 200504


\bibitem{channelASHolevo} Holevo A S 2007 {\it Probl. Inf. Transm.} {\bf 43} 1



\bibitem{het} Weedbrook C, Lance A M, Bowen W P, Symul T, Ralph T C and Lam P K 2004 {\it Phys. Rev. Lett.} {\bf 93} 170504



\bibitem{TSymulexperiment} Symul T, Alton D J, Assad S M, Lance A M, Weedbrook C, Ralph T C and Lam P K 2007 {\it Phys. Rev. A} {\bf 76} 030303


\end{thebibliography}
\end{document}